\documentclass[final,1p,times]{elsarticle}

\usepackage{amssymb}
\usepackage{graphicx}
\usepackage{amsmath}

\usepackage{amsthm}

\usepackage{lineno}

\journal{Journal Astronomy and Space Science}

\begin{document}

\begin{frontmatter}



\title{On the existence of regular tetrahedral non-homothetic homographic solution}

\author{Jaewoo Kim         }
\ead{macqueen01@hafs.hs.kr}
\address{Hankuk Academy of Foreign Studies, Yongin 17035, Korea}





\begin{abstract}
It is well known that the three-body problem has few analytical solutions in certain symmetrical constraints; the Lagrangian triangular solution is one of them. This triangular solution has been revisited by R.Broucke and H.Lass in 1971, concerning three relative position vectors pointing from one mass to another. This paper proposes a significant advance to the method, extended to four arbitrary masses on the vertices of a tetrahedron. The research provides a geometrical proof that under such constraint, only homothetic solution is possible which agrees with the conclusion brought by the article 371 of Wintner (1941).

\end{abstract}

\begin{keyword}
Four-body problem \sep Tetrahedral solution \sep Celestial mechanics \sep N-body problem


\end{keyword}

\end{frontmatter}


\section{INTRODUCTIONS}

Before Poincare’s concrete math which proved that differential equations issued by numeral masses (more than two masses) do not yield a general solution different from that of two masses, some mathematicians and physicists had proposed few particular solutions with several different constraints: certain mass combinations, orbits, or initial velocity (Bernat et al. 2009, Delgado and  Vidal 1999). It is also well known that Euler and Lagrange provided the collinear solution and the triangular solution to $ i=3 $ case each (Hestenes 1986), where $ i $ indicates the number of masses. Their abstract solutions have been revisited by R.Broucke and H.Lass with scrupulous mathematical manipulation in vector calculation which Lagrange and Euler had not yet experienced (Broucke and Lass 1973). They introduced their newest Lagrangian with relative vectors, rather than general position vectors, which further allowed them to decouple each mass’s force vectors by posing certain constraints, thus making perturbation vector $\Omega$ zero. This mathematical manipulation on Newtonian motion equation, in similar suit, if applied to the four body case, yields some familiar decoupled equations that previously discovered in R.Broucke’s study. Only equilateral tetrahedral configuration acts as constraint in the equation, and, alongside with several other three body solutions (Montgomery 2001, Hsiang and Straume 2008, Murnaghan 1936), have been a long question of the n-body problem. R. Lehmann-Filhes first brought this configuration in which four arbitrary masses directly fall toward the center of mass, which nowadays is referred to as homothetic solution (Lehmann-Filhes 1891). With further notion in (Wintner 1941), it has been proven that except for some obvious cases, placing equal $(i-1)$ masses on every corners of $(i-1)$-gon and one arbitrary mass in the middle, it is possible to construct homothetic solutions only for a finite set of particular values of $i$. This intuitive, yet imperfect, solution further left the question whether this configuration could extend to non-collinear non-flat solution, non-homothetic homographic solution. It is found that non-homothetic homographic solution of $i$-gon doesn't exist for any integer $i$ larger than four as it is proven in §371 of Wintner (1941). This paper offers another straightforward and elementary approach to the conclusion brought by Wintner (1941).

\section{The Equations of Motion in Barycentric Coordinates}
Let $ \vec{x_{1}} $, $ \vec{x_{2}} $, $ \vec{x_{3}} $, and $ \vec{x_{4}} $ represent the position vectors starting from the center of mass of the four point-masses $ m_{1} $, $ m_{2} $, $ m_{3} $, and $ m_{4} $ in the barycentric rectangular coordinate with three dimensions. Under Newtonian theory of Gravity, each four point-masses pull each other with the force represented in the following:\\
\begin{equation}
\begin{aligned}
\ddot{\vec{x}}_{1}=-Gm_{2}\frac{\vec{x}_{1}-\vec{x}_{2}}{{\left \|\vec{x}_{1}-\vec{x}_{2} \right \|}^{3}}-Gm_{3}\frac{\vec{x}_{1}-\vec{x}_{3}}{{\left \|\vec{x}_{1}-\vec{x}_{3} \right \|}^{3}}-Gm_{4}\frac{\vec{x}_{1}-\vec{x}_{4}}{{\left \|\vec{x}_{1}-\vec{x}_{4} \right \|}^{3}},\\
\ddot{\vec{x}}_{2}=-Gm_{1}\frac{\vec{x}_{2}-\vec{x}_{1}}{{\left \|\vec{x}_{2}-\vec{x}_{1} \right \|}^{3}}-Gm_{3}\frac{\vec{x}_{2}-\vec{x}_{3}}{{\left \|\vec{x}_{2}-\vec{x}_{3} \right \|}^{3}}-Gm_{4}\frac{\vec{x}_{2}-\vec{x}_{4}}{{\left \|\vec{x}_{2}-\vec{x}_{4} \right \|}^{3}},\\
\ddot{\vec{x}}_{3}=-Gm_{1}\frac{\vec{x}_{3}-\vec{x}_{1}}{{\left \|\vec{x}_{3}-\vec{x}_{1} \right \|}^{3}}-Gm_{2}\frac{\vec{x}_{3}-\vec{x}_{2}}{{\left \|\vec{x}_{3}-\vec{x}_{2} \right \|}^{3}}-Gm_{4}\frac{\vec{x}_{3}-\vec{x}_{4}}{{\left \|\vec{x}_{3}-\vec{x}_{4} \right \|}^{3}},\\
\ddot{\vec{x}}_{4}=-Gm_{1}\frac{\vec{x}_{4}-\vec{x}_{1}}{{\left \|\vec{x}_{4}-\vec{x}_{1} \right \|}^{3}}-Gm_{2}\frac{\vec{x}_{4}-\vec{x}_{2}}{{\left \|\vec{x}_{4}-\vec{x}_{2} \right \|}^{3}}-Gm_{3}\frac{\vec{x}_{4}-\vec{x}_{3}}{{\left \|\vec{x}_{4}-\vec{x}_{3} \right \|}^{3}}.
\end{aligned}
\end{equation}
\section{The Equations of Motion in Relative Coordinates
}
\begin{figure}
	\includegraphics[scale=0.7]{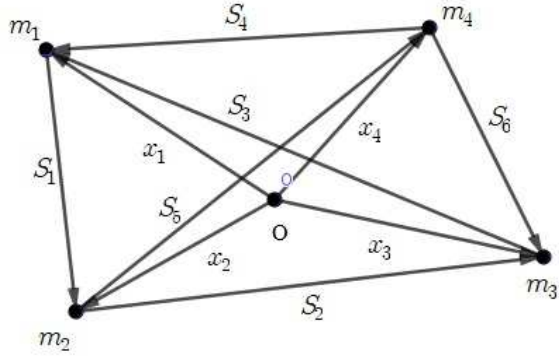}
\caption{Four point-masses and their barycentric, and relative position vectors are introduced.}	
\label{fig:1}
\end{figure}

To check if the four point-masses in the system experience a symmetrical force directed to the center of the mass, the relative forces acted upon each mass should be indicated by the six relative position vectors, $ S_{n} $ (see Fig.~\ref{fig:1}): 
\begin{equation}
\begin{aligned}
S_{1}=\vec{x}_{2}-\vec{x}_{1},\quad S_{4}=\vec{x}_{4}-\vec{x}_{1},\\
S_{2}=\vec{x}_{3}-\vec{x}_{2},\quad S_{5}=\vec{x}_{4}-\vec{x}_{2},\\
S_{4}=\vec{x}_{1}-\vec{x}_{3},\quad S_{6}=\vec{x}_{3}-\vec{x}_{4},
\end{aligned}
\end{equation}\\
\begin{equation}
\begin{aligned}
S_{1}+S_{2}+S_{3}=0,\\
S_{1}+S_{4}+S_{5}=0,\\
S_{5}+S_{2}-S_{6}=0,
\end{aligned}
\end{equation}\\
\begin{equation}
\begin{aligned}
m_{1}\vec{x}_{1}+m_{2}\vec{x}_{2}+m_{3}\vec{x}_{3}+m_{4}\vec{x}_{4}=0
\end{aligned}.
\end{equation}\\
With simple substitution of the equation (1) introduced earlier in section 2 in (2), we get a complex system of six equations of motion in the relative frame. A certain constraint will be introduced in the following paragraph that reduces the six rigorous equations into six simple equations.

	Here, one should note that in the three-dimensional rectangular coordinate, four point-masses are allowed to form a tetrahedron with six edges. If the masses take place at the vertices of an equilateral tetrahedron with six edges with the same length, the equations of relative motion can be reduced to the following after a little algebra:\\
\begin{equation}
\begin{aligned}
\ddot{S}_{n}=-G\frac{\mu S_{n}}{{\left \| S_{n} \right \|}^{3}}\\
\mu =m_{1}+m_{2}+m_{3}+m_{4}.
\end{aligned}
\end{equation}\\
Broucke and Lass arrive at this similar conclusion in the $ n=3 $ case. Just by eliminating one mass in our equation and keeping everything the same, the following three symmetric equations of relative motion and a perturbation vector are induced:\\
\begin{equation}
\begin{aligned}
\ddot{S}_{n}=-G\frac{\mu S_{n}}{{\left \| S_{n} \right \|}^{3}}+m_{n}\Omega ,
\end{aligned}
\end{equation}\\
\begin{equation}
\begin{aligned}
\Omega =G\left ( \frac{S_{1}}{{\left \| S_{1} \right \|}^{3}}+\frac{S_{2}}{{\left \| S_{2} \right \|}^{3}}+\frac{S_{3}}{{\left \| S_{3} \right \|}^{3}} \right ).
\end{aligned}
\end{equation}\\
In (6), we readily see same $\Omega$ multiplied with $m_{n}$ at the tail of each $ \ddot{S}_{1} $, $ \ddot{S}_{2} $, and $ \ddot{S}_{3} $. Some certain constraints that set $ \Omega=0 $ are the key to the particular solution to the three-body problem, and $ \Omega $ itself is a perturbation vector whose magnitude affects the stability of the orbit of the particular solution. This perturbation vector also appears in the $ n=6 $ case where its increase in number hints us analytic complexity of the system of the general four-body problem which results in chaos in the future phase. Even without numerical integration, by qualitatively probing these vectors, one might predict the tendency of each mass’s phase shortly after the initial condition.

	By giving the constraint suggested above —equilateral tetrahedron— to the equations of relative motion, we can decouple the equations, and reduce it into a simple two-body problem, which can be afforded with familiar analytic methods. 

\section{Force Directed toward the Center of Mass}
We've seen under what condition the equations would decouple. The consequence of the only possible constraint in three-dimensional space —$ S_{1}^{2}=S_{2}^{2}=...=S_{5}^{2}=S_{6}^{2} $— is simply just four point-masses pulling each other to the center of mass during their period. From this, it is qualitatively deducible that the masses would converge, or sometimes diverge, to and from the center of mass, staying in the same shape —equilateral tetrahedron— only varying in its size and orientation. This conclusion may hint that each mass experiences Newtonian central force directed to the center of four point-masses, forming a conic orbit around the barycentre, and this section provides a mathematical foundation to it.

	We start by looking for a direct relationship between barycentric coordinate vectors —$ \vec{x}_{i} $— and relative coordinate vectors —$ S_{n} $—. Solving (2) and (4) for , we get (8):\\
\begin{equation}
\begin{aligned}
m_{3}S_{3}-m_{2}S_{1}-m_{4}S_{4}=\mu \vec{x}_{1},\\
m_{1}S_{1}-m_{4}S_{4}-m_{3}S_{2}=\mu \vec{x}_{2},\\
m_{4}S_{6}-m_{1}S_{3}-m_{2}S_{2}=\mu \vec{x}_{3},\\
m_{2}S_{5}-m_{3}S_{6}-m_{1}S_{4}=\mu \vec{x}_{4}.
\end{aligned}
\end{equation}\\
Note that the tetrahedron varies its size and orientation in three-dimensional space. Unlike the conventional triangular solution of the three-body problem, masses have three general coordinates and six degrees of freedom. Thus, we employ quaternion to express each $ S_{n} $ in respect of $ S_{1} $.

\begin{figure}
	\includegraphics[scale=0.5]{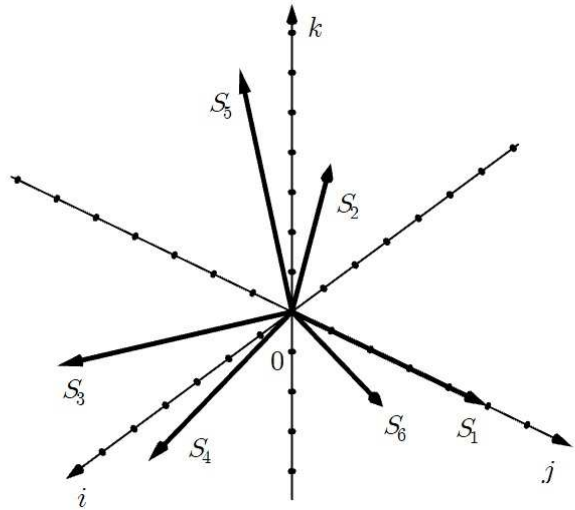}
\caption{All relative position vectors are shifted to the origin and are expressed with quaternions in (9).}	
\label{fig:2} 
\end{figure}

	First, recall that a vector rotation made in a certain plane with its bi-vector, $ \hat{n} $, is referred to as
\begin{equation}\notag
{\vec{v}}'=\vec{v}\times \left [ \cos \theta +\sin \theta \cdot \hat{n} \right ]
\end{equation}\\
where $ \theta $ denotes $ {\vec{v}}' $’s angle of rotation respect to $ \hat{n} $. With this handy formula (Wintner 1947), one could derive the following after some straightforward calculations involving geometries (see Fig.~\ref{fig:2}):\\
\begin{flalign}
S_{2}&=S_{1}\times \left ( -\frac{1}{2}+\frac{\sqrt{3}}{2}\hat{k} \right ),\quad S_{3}=S_{4}\times \left ( -\frac{1}{2}-\frac{\sqrt{3}}{2}\hat{k} \right ),&&\\\nonumber
S_{4}&=S_{1}\times \left ( -\frac{1}{2}+\frac{\sqrt{6}}{3}\hat{i}-\frac{\sqrt{3}}{6}\hat{k} \right ),\quad S_{5}=S_{1}\times \left ( -\frac{1}{2}-\frac{\sqrt{6}}{3}\hat{i}+\frac{\sqrt{3}}{6}\hat{k} \right ),&&\\\nonumber
S_{6}&=S_{1}\times \left ( -\frac{\sqrt{2}}{3}\hat{i}+\frac{\sqrt{7}}{3}\hat{k} \right ).&&
\end{flalign}\\
Relating (9) and (8), barycentric vectors then can be written in the form (10):\\
\begin{align}
\mu \vec{x}_{1}&=S_{1}\times \left [ \frac{m_{4}-m_{3}}{2}-m_{2}-\frac{\sqrt{6}m_{4}}{3}\hat{i}+\left ( \frac{\sqrt{3}m_{4}-3\sqrt{3}m_{3}}{6} \right )\hat{k} \right ],\\\nonumber
\mu \vec{x}_{2}&=S_{1}\times \left [ \frac{m_{3}+m_{4}}{2}+m_{1}+\frac{\sqrt{6}m_{4}}{3}\hat{i}-\left ( \frac{\sqrt{3}m_{4}+3\sqrt{3}m_{3}}{6} \right )\hat{k} \right ],\\\nonumber
\mu \vec{x}_{3}&=S_{1}\times \left [ \frac{m_{1}+m_{2}}{2}-\frac{\sqrt{2}m_{4}}{3}\hat{i}+\left ( \frac{2\sqrt{7}m_{4}+3\sqrt{3}m_{1}-3\sqrt{3}m_{2}}{6} \right )\hat{k} \right ],\\\nonumber
\mu \vec{x}_{4}&=S_{1}\times \left [ \frac{m_{1}-m_{2}}{2}+\left ( \frac{\sqrt{2}m_{3}-\sqrt{6}m_{2}-\sqrt{6}m_{1}}{3} \right )\hat{i}\right ]\\\nonumber
&\qquad \qquad \qquad \qquad \qquad \quad +S_{1}\times \left [ \left ( \frac{\sqrt{3}m_{2}+\sqrt{3}m_{1}-2\sqrt{7}m_{3}}{6} \right )\hat{k} \right ].
\end{align}\\
The terms multiplied after $ S_{n} $ are linear transformation operators that give rotations to $ S_{1} $ in three-dimensional space. Since they are independent from time $ t $, the following relation between acceleration vectors of barycentric coordinate and relative coordinate are induced after differentiating (10) twice:\\
\begin{flalign}
\ddot{\vec{x}}_{1}&=\dfrac{\ddot{S_{1}}}{\mu}\times \left [ \frac{m_{4}-m_{3}}{2}-m_{2}-\frac{\sqrt{6}m_{4}}{3}\hat{i}+\left ( \frac{\sqrt{3}m_{4}-3\sqrt{3}m_{3}}{6} \right )\hat{k} \right ],&&\\\nonumber
\ddot{\vec{x}}_{2}&=\dfrac{\ddot{S_{1}}}{\mu}\times \left [ \frac{m_{3}+m_{4}}{2}+m_{1}+\frac{\sqrt{6}m_{4}}{3}\hat{i}-\left ( \frac{\sqrt{3}m_{4}+3\sqrt{3}m_{3}}{6} \right )\hat{k} \right ],&&\\\nonumber
\ddot{\vec{x}}_{3}&=\dfrac{\ddot{S_{1}}}{\mu}\times \left [ \frac{m_{1}+m_{2}}{2}-\frac{\sqrt{2}m_{4}}{3}\hat{i}+\left ( \frac{2\sqrt{7}m_{4}+3\sqrt{3}m_{1}-3\sqrt{3}m_{2}}{6} \right )\hat{k} \right ],&&\\\nonumber
\ddot{\vec{x}}_{4}&=\dfrac{\ddot{S_{1}}}{\mu}\times \left [ \frac{m_{1}-m_{2}}{2}+\left ( \frac{\sqrt{2}m_{3}-\sqrt{6}m_{2}-\sqrt{6}m_{1}}{3} \right )\hat{i}\right ]\\\nonumber
&\qquad \qquad \qquad \qquad \qquad \quad +\dfrac{\ddot{S_{1}}}{\mu}\times \left [ \left ( \frac{\sqrt{3}m_{2}+\sqrt{3}m_{1}-2\sqrt{7}m_{3}}{6} \right )\hat{k} \right ].&&
\end{flalign}
Finally, by substituting (5) in (11) and again in (10), we get (12) and readily notice these motion equations for the four-body problem regress to familiar two-body problem:\\
\begin{equation}
\ddot{\vec{x}}_{i}=-G\frac{\mu \vec{x}_{i}}{{\left \| S_{1} \right \|}^{3}}.
\end{equation}\\
To see if the masses would experience the inverse square law, we start by calculating the length of $S_1$ and $\vec{x}_i$ to substitute them with equation (12). From equation (10), one can easily relate their length as follows.
\begin{equation}
\left \| S_{1} \right \|=\frac{\mu \left \| \vec{x}_{i} \right \|}{\kappa_{i} ^{\frac{1}{2}}},
\end{equation}
\begin{equation}
\begin{aligned}
\kappa _{1}=m_{2}^{2}+m_{3}^{2}+m_{4}^{2}+m_{1}m_{3}+m_{2}m_{4}+m_{3}m_{4},\\
\kappa _{2}=m_{1}^{2}+m_{3}^{2}+m_{4}^{2}+m_{1}m_{3}+m_{1}m_{4}+m_{3}m_{4},\\
\kappa _{3}=m_{1}^{2}+m_{2}^{2}+m_{4}^{2}+m_{1}m_{2}+m_{1}m_{4}+m_{2}m_{4},\\
\kappa _{4}=m_{1}^{2}+m_{2}^{2}+m_{3}^{2}+m_{1}m_{2}+m_{1}m_{3}+m_{2}m_{3}.
\end{aligned}
\end{equation}
By substituting (13) in (12), we notice the masses experience Newtonian Gravitational force: 
\begin{equation}
\ddot{\vec{x}}_{i}=-G\frac{\kappa _{i}^{\frac{3}{2}}\vec{x}_{i}}{\mu ^{2}\left \| \vec{x}_{i} \right \|^{3}}.
\end{equation}
\section{The Plane Solution}
In section 4, we’ve proved that the masses experience central force while they form an equilateral tetrahedral configuration. This means, in other words, if the configuration breaks, they will no longer be treated as a particular solution. We now prove that there is no homographic solution which is not homothetic by contradiction. If there is such solution, we have to meet more conditions with which the masses still keep their tetrahedral configuration while they orbit. Note that the orbits must be conic since masses experience the central force directed to the center of mass within their equilateral tetrahedral configuration. Also, note that conic describes the motion of a body constrained onto a plane, only affected by the central force on the same plane. Thus, if non-planer homographic solution, which is non-homothetic homographic solution, exists each masses should be placed on four different planes with an independent bivector. 

\begin{figure}
	\includegraphics[scale=0.6]{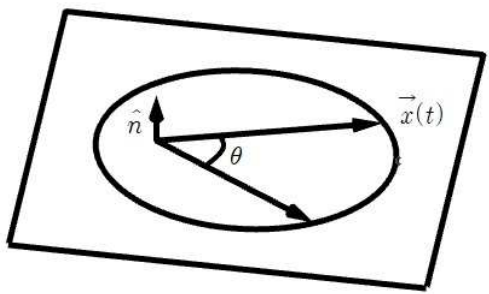}
\caption{$ \vec{x}(t) $ is a set of vector rotations of $ \vec{r}(t) $ which only varies in size through angle $ \theta $.}	
\label{fig:3} 
\end{figure}

	Recall how we expressed a vector rotation respect to a certain bi-vector in section 4. We use it again to describe the motion of a mass under the central force on a plane with a given bi-vector, $ \hat{n} $. It is familiar using the polar coordinate to analyse the motion of a mass under the central force. In the system appears in (Marion and Jerry 2013), we introduce $ r(t) $ and $ \theta(t) $, each indicating the distance between the mass and the source of the central force and the angle from the mass’s initial state measured in time $ t $. Here, based on such familiar qualities of the polar coordinate system and the vector rotation mentioned earlier in the paragraph, the motion equation of a mass on a certain constraint plane is introduced:\\
\begin{equation}
\vec{x}(t)=\vec{r}(t)\times \left [ \cos \theta \left ( t \right )+\sin \theta \left ( t \right )\cdot \hat{n} \right ].
\end{equation}\\
$ \vec{x}(t) $ is a set of vector rotations of $ \vec{r}(t) $ over time $ t $ (see Fig.~\ref{fig:3}).\\

Note that the four masses orbit in the same angular velocity and period. This assumption is essential for non-planer homographic solutions, since there is no possible way the masses keeping their configuration without the synchronized period of each conic. Upon these, we construct the following four novel barycentric position vectors —(17)— varying through time.\\
\begin{equation}
\begin{aligned}
\vec{x}_{1}(t)=\vec{r}_{1}(t)\times \left [ \cos \theta \left ( t \right )+\sin \theta \left ( t \right )\cdot \hat{n}_{1} \right ],\\
\vec{x}_{2}(t)=\vec{r}_{2}(t)\times \left [ \cos \theta \left ( t \right )+\sin \theta \left ( t \right )\cdot \hat{n}_{2} \right ],\\
\vec{x}_{3}(t)=\vec{r}_{3}(t)\times \left [ \cos \theta \left ( t \right )+\sin \theta \left ( t \right )\cdot \hat{n}_{3} \right ],\\
\vec{x}_{4}(t)=\vec{r}_{4}(t)\times \left [ \cos \theta \left ( t \right )+\sin \theta \left ( t \right )\cdot \hat{n}_{4} \right ].
\end{aligned}
\end{equation}\\
Recall that masses should keep an equilateral tetrahedral configuration during their procession, constrained by equation (4). Substitution of (17) in (4) results to the following:\\
\begin{equation}
\sum_{i=1}^{4}m_{i}\vec{x}_{i}\left ( t \right )=\cos \theta \left ( t \right )\sum_{i=1}^{4}m_{i}\vec{r}_{i}\left ( t \right )+\sin \theta \left ( t \right )\sum_{i=1}^{4}\left ( m_{i}\vec{r}_{i}\left ( t \right )\times \hat{n}_{i} \right )=0.
\end{equation}\\
Trivially, the summation where cosine is multiplied becomes zero, and (18) is again reduced to
\begin{equation}
\sum_{i=1}^{4}\left ( m_{i}\vec{r}_{i}\left ( t \right )\times \hat{n}_{i} \right )=0
\end{equation}
or
\begin{equation}
\sin{\theta\left(t\right)}=0,
\end{equation}
where (19) and (20) imply respectively non-homothetic homographic solution and homothetic solution. We further notice that one of (19) and (20) must be true by definition of (18), and this lemma agrees, or provides another proof, to the article 371 of Wintner (1941) that every homographic solution is either planar or homothetic but may be both.

\begin{figure}
	\includegraphics[scale=0.6]{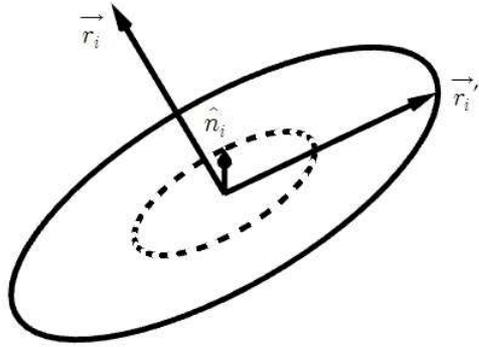}
\caption{Both $ {\vec{r}_{i}}' $ and $ \hat{n}_{i} $ are constrained in the plane on which they make concentric circles about the original position vector, $ \vec{r}_{i} $.}	
\label{fig:4} 
\end{figure}

	As the orbits each mass draws should share the same eccentricity —the ratio of the distances between the center and each mass is conserved during a period— and as the directions of $ \vec{r}_{i}(t) $ are solid, we can further reduce our concern on (19)’s time dependence. Thus, we simply take an arbitrary $ \vec{r}_{i}(t) $, which will be $ \vec{r}_{i}(0) $ in the paper, and refer to it as $\vec{r}_{i}$. With this in mind, (19) can be looked upon as an equilateral tetrahedron in a different orientation: each position vector rotated 90 degrees respect to their bivectors, conserving the ratio between the lengths of $\vec{r}_{i}$. The rotated position vectors will be indicated as ${\vec{r}_{i}}'$. Though both $\vec{r}_{i}$ and $\hat{n}_{i}$ are unknown quantities, we readily realize they are perpendicular to each other satisfying $ {\vec{r}_{i}}'\perp \hat{n}_{i}\perp {\vec{r}_{i}}' $. What the formulation suggests is the configuration of the two vectors —${\vec{r}_{i}}'$ and $\hat{n}_{i}$— making two concentric circles while rotating, constrained on a plane with its normal vector $\vec{r}_{i}$ (see Fig.~\ref{fig:4}).

	Our goal is to find the right pair of ${\vec{r}_{i}}'$ which \textit{resembles} the pair of $ \vec{r}_{i} $, each $ {\vec{r}_{i}}' $ perpendicular to $ \vec{r}_{i} $. The term \textit{resemblance} is here used to indicate the masses’ equilateral tetrahedral configuration. In other words, no change should occur in the relative coordinate in respect of $ {\vec{r}_{i}}' $. Therefore, the relation established in the pair of $ \vec{r}_{i} $ should be preserved in the pair of ${\vec{r}_{i}}'$ in their relative coordinate. Notice what this \textit{resemblance} suggests us: the angles ${\vec{r}_{i}}'$ make with each other equal the angles $ \vec{r}_{i} $ make with each other. Thus,\\
\begin{equation}
\begin{aligned}
{\vec{r}_{1}}'\cdot {\vec{r}_{2}}'=\vec{r}_{1}\cdot \vec{r}_{2},\quad {\vec{r}_{3}}'\cdot {\vec{r}_{4}}'=\vec{r}_{3}\cdot \vec{r}_{4},\\
{\vec{r}_{2}}'\cdot {\vec{r}_{3}}'=\vec{r}_{2}\cdot \vec{r}_{3},\quad {\vec{r}_{4}}'\cdot {\vec{r}_{1}}'=\vec{r}_{4}\cdot \vec{r}_{1}
\end{aligned}
\end{equation}\\
should hold. Take an arbitrary pair of $\vec{R}_{i}$ we can randomly assign, whose components are equal in length and are  perpendicular to those of $ \vec{r}_{i} $. Let the angle $\vec{R}_{i}$ makes with ${\vec{r}_{i}}'$ a scalar value $ \theta_{i} $. Then, ${\vec{r}_{i}}'$ can be looked upon as a result of the vector rotation of an arbitrary $\vec{R}_{i}$ about $ \vec{r}_{i} $:\\
\begin{equation}
{\vec{r}_{i}}'=\vec{R}_{i}\times \left [ \cos \theta _{i}+\sin \theta_{i} \cdot \frac{\vec{r}_{i}}{\left \| \vec{r}_{i} \right \|} \right ].
\end{equation}\\
By relating (17) and (18), we get the following:\\
\begin{equation}
A_{ij}\cos \theta _{i}\cos \theta _{j}+B_{ij}\sin \theta _{i}\sin \theta _{j}+C_{ij}\cos \theta _{i}\sin \theta _{j}+D_{ij}\sin \theta _{i}\cos \theta _{j}=\vec{r}_{i}\cdot \vec{r}_{j},
\end{equation}
\begin{equation}
\begin{aligned}
A_{ij}=\vec{R}_{i}\cdot \vec{R}_{j},\quad \quad C_{ij}=\vec{R}_{i}\cdot \left ( \frac{\vec{R}_{j}\times \vec{r}_{j}}{\left \| \vec{r}_{j} \right \|} \right ),\\
B_{ij}=\frac{\left ( \vec{R}_{i}\times \vec{r}_{i} \right )\cdot \left ( \vec{R}_{j}\times \vec{r}_{j} \right )}{\left \| \vec{r}_{i} \right \|\left \| \vec{r}_{j} \right \|}, \quad D_{ij}=\vec{R}_{j}\cdot \left ( \frac{\vec{R}_{i}\times \vec{r}_{i}}{\left \| \vec{r}_{i} \right \|} \right ).
\end{aligned}
\end{equation}\\
Note that $ A_{ij} $, $ B_{ij} $, $ C_{ij} $, $ D_{ij} $, and $\vec{r}_{i}\cdot \vec{r}_{j}$ are constants one can easily induce through straightforward calculations based on the vector component values of $\vec{R}_{i}$ we assigned earlier. Consider that $i$ and $j$ both vary from 0 to 4. We have twelve equations in maximum and four equations in minimum with four unknown variables —$ \theta_{1} $, $ \theta_{2} $, $ \theta_{3} $, and $ \theta_{4} $. As these equations are solved, and if they yield real solutions for $ \theta_{i} $, it is possible to induce $\vec{x}_{i}(t)$ from (17), as we get ${\vec{r}_{i}}'$ from the solutions:\\
\begin{equation}
\vec{x}_{i}(t)= \frac{\left \| \vec{r}_{i}(t) \right \|}{\left \| \vec{r}_{i} \right \|}\left [ \vec{r}_{i}\cos \theta (t)+{\vec{r}_{i}}'\sin \theta (t) \right ].
\end{equation}\\
(25) can be rewritten in a practical form in terms of $ \theta $ and $ \varepsilon $, taking $ \vec{r}_{i} $ as perihelion:\\
\begin{equation}
\vec{x}_{i}(\theta )= \frac{1-\varepsilon }{1-\varepsilon \cos \theta }\left [ \vec{r}_{i}\cos \theta +{\vec{r}_{i}}'\sin \theta \right ]
\end{equation}\\
where $\varepsilon$ is the eccentricity shared by all four orbits which can be chosen arbitrarily.

But let me remind that these resulting trajectories-(25) and (26)- only make sense if and only if the equations in (23) yield real solutions for all $ \theta_{i} $. That is, if this result is numerically solved and is denied, (19) no longer holds, and we are to take (20) to be true, which agrees to the conclusion of the article 371 of Wintner (1941). By taking simple but long numerical approach, taking an arbitrary pair of vectors $\vec{R}_{i}$ each with equal in length and perpendicular to the pair of $ \vec{r}_{i} $ and substituting them in (24) and again in (23), one can get solutions for $\cos \theta _{i}$ and $\sin \theta _{i}$. Note that even a single numerical result proves or disproves the validity of (19). In this research, numerical calculation proved that minimun at one of $\cos \theta _{i}$ and $\sin \theta _{i}$ had to exceeded 1, which implies $ \theta_{i} $ involve complex values. This, again, is enough to reject (19) and conclude that (20) holds. That is, only homothetic central tetrahedral configuration solution exists for the four-body problem which agrees with the article 371 of Wintner (1941).
 
\section{FIGURES AND TABLES }

\end{document}